# Ancilla-free Reversible Logic Synthesis via Sorting

Anupam Chattopadhyay[1] and Sharif MD Khairul Hossain[2]

[1] School of Computer Science and Engineering,
Nanyang Technological University, Singapore
[2] Department of Computer Science,
RWTH Aachen University, Germany

**Abstract.** Reversible logic synthesis is emerging as a major research component for post-CMOS computing devices, in particular Quantum computing. In this work, we link the reversible logic synthesis problem to sorting algorithms. Based on our analysis, an alternative derivation of the worst-case complexity of generated reversible circuits is provided. Furthermore, a novel column-wise reversible logic synthesis method, termed **RevCol**, is designed with inspiration from radix sort. Extending the principles of RevCol, we present a hybrid reversible logic synthesis framework. The theoretical and experimental results are presented. The results are extensively benchmarked with state-of-the-art ancilla-free reversible logic synthesis methods.

## 1 Introduction

Physically reversible computation is an integral part of Quantum computing. The performance breakthrough in several Quantum algorithms [1] compared to their classical counterpart as well as the hype around practical Quantum computers [2] ushered in the wave of reversible computing. In a major theoretical development [3], it is shown that physical reversibility must be accompanied by logical reversibility. Current band of logical primitives used in charge-based computing (e.g. Boolean NAND operation) is often irreversible and hence, unusable for logically reversible computation. Consequently, major research attention is given towards the synthesis of a Boolean function using a set of reversible logic gates, a problem otherwise known as reversible logic synthesis. Practical experiments with reversible logic gates is continuously driving the research to synthesize a reversible circuit with optimized performance.

### 1.1 Preliminaries

An $n$-variable Boolean function $f$ is a mapping $f : GF(2^n) \to GF(2)$. Another representation is a mapping $f : \{0,1\}^n \to \{0,1\}$, which is known as the truth table representation. Using any basis of $GF(2^n)$, we can express each $x \in GF(2^n)$ as an $n$-tuple $(x_1 x_2 \dots x_n)$, $x_i \in GF(2)$, $i = 1, \dots, n$.

An $n$-variable Boolean function is *reversible* if all its output patterns map uniquely to an input pattern and vice-versa. It can be expressed as an $n$-input, $n$-output bijection or alternatively, as a permutation function over the truth value set $\{0, 1, \dots 2^{n-1}\}$. The problem of reversible logic synthesis is to map such a reversible Boolean function on a reversible logic gate library. Prominent reversible logic gates include NOT, Feynman (or CNOT), Toffoli (or CCNOT), Fredkin, Generalized Toffoli ($Tof_n$), and Generalized Fredkin ($Fred_n$) [4].

When additional input Boolean variables are needed for constructing the output function, those are referred as *ancilla*. For a given irreversible Boolean function, the minimum ancilla count can be exactly derived. For practical implementation purposes, it is desirable to limit the ancilla count to this minimum value. However, the synthesis methods often introduce additional ancilla lines to perform trade-off with other performance objectives, e.g. Quantum cost (QC). QC of a reversible gate is its implementation cost in Quantum technologies [5]. A few synthesis methods guarantee that the ancilla count is restricted to the minimum number. These are known as ancilla-free reversible logic synthesis methods.

Logical depth is another performance objective. In a reversible circuit, A level is defined as a sub-sequence of elementary gates that can be applicable in parallel [6]. The number of logical levels in a circuit is called logical depth. A decrease in logical depth reduces of the execution time of a circuit and it mitigates the effect of decoherence [7].

## 2 Related Work and Motivation

During the last decade, in the field of reversible logic there has been several synthesis techniques, which derived optimal reversible circuits in terms of gate counts and QC [8–10]. Scalability of these optimal solutions to arbitrary Boolean functions for large number of variables remains an interesting open problem, which prompted researchers to take multiple research routes.

In one direction, scalability is the prime goal and incurring ancilla overhead is allowed. These methods borrowed heavily from classical logic synthesis techniques like Binary Decision Diagrams [11] or Exclusive Sum-of-Product formulation [12, 13]. Here, we restrict our discussion to ancilla-free methods. A detailed discussion on various reversible logic synthesis techniques can be found in [14].

A general technique of reversible logic synthesis, that is directly applied for truth-table representation of Boolean function is presented in [15]. This technique, referred as MMD, remained a prominent, ancilla-free synthesis method, due to its simple algorithmic structure and excellent performance compared to optimal results of 3 and 4 variable circuits.

Several important previous works exist relating sorting and ancilla-free reversible logic synthesis. The relation between reversible logic synthesis and a permutation group has been established in [16]. Later, in [17] and [18], swapping elements of a permutation $\pi$ is used as a technique to achieve the Identity permutation $\mathbb{I}$. Intuitively, $\pi$ and $\mathbb{I}$ are the output and input of the reversible function respectively. Interestingly both the works [17, 18] restricted the swaps to the bitstrings with a Hamming distance of 1, which could be realized with a $Tof_n$ gate. The connection to the general class of sorting algorithm is not explored. None of these works could improve upon the MMD, in terms of gate count. In [16], it has been shown that any reversible Boolean function, when expressible using even permutation, can be realized using only $Tof_0$ and $Tof_1$ gates. This is accomplished by first, decomposing the even permutation into a series of disjoint 3-cycles and then mapping these 3-cycles to reversible gates. Corresponding to this implementation flow, the worst-case circuit complexity is also derived. A k-cycle-based synthesis method for reversible functions was proposed in [19]. A decomposition algorithm was proposed to decompose a large cycle into a set of elementary cycles. The synthesis algorithm used this a set of elementary cycles to construct circuits.

**Motivation:** Despite the rich body of research in sorting and permutation decomposition, its connection to reversible logic synthesis is explored little. In this paper, we attempt a connection between various sorting algorithms and reversible logic synthesis. We present theoretical results and experimental evidence to demonstrate improved results compared to state-of-the-art ancilla-free reversible logic synthesis.

The rest of this paper is organized as following. In section 3, the connection between reversible logic synthesis and sorting is established. The challenges in a sorting-based reversible logic synthesis flow are described in detail. We present a practical algorithm for reversible logic synthesis based on radix sort in section 4 and present several optimizations within the scope of this algorithm. Experimental results are presented in section 5. The paper is summarized with an outline of future works in section 6.

## 3 Reversible Logic Synthesis and Sorting

A reversible Boolean function can be defined as an ordered set of integers corresponding to the a permutation of its domain. Hence, the reversible circuit, when traversed from output towards input, essentially converts the permutation to the Identity specification.

**Definition 1.** *Let $S$ be an arbitrary nonempty set. A bijection (a one-to-one, onto mapping) of $S$ onto itself is called a permutation of $S$.*

**Definition 2.** *Given a function $f : [0, n] \rightarrow [0, n]$, the functional digraph $G(f) = (V, E)$ associated with $f$ is a directed graph with $V = \{0, ..., n\}$ and $E = \{(v, f(v)) for\ each\ v \in V\}$*

For a permutation $\pi$ of $0, 1, \cdots, n$, G is a collection of disjoint cycles. For each cycle $c = (a_1, a_2, \cdots, a_k)$, the permutation cyclically shifts all entries in $c$ and keeps all other elements fixed.

$$a_1 \rightarrow a_2 \rightarrow a_3 \rightarrow \cdots a_k \rightarrow a_1$$

A cycle of length 2 is called a transposition. A (transposition) decomposition $\sigma$ of a permutation $\pi$ is a sequence $t_m, \cdots, t_1$ of transpositions $t_i$ whose product is $\pi$. A sorting $s$ of a permutation $\pi$ is a sequence of transpositions that transform $\pi$ into $\mathbb{I}$. In other words, $s \cdot \pi = \mathbb{I}$.

The following lemma quantifies the impact of sorting by using transpositions on reversible logic synthesis.

**Lemma 1.** *Given a permutation $\pi$, any 2 elements $\pi_i$ and $\pi_j$ can be swapped using reversible gates. The cost of this operation is $2 \cdot \delta(\pi_i, \pi_j) - 1$, where $\delta(\pi_i, \pi_j)$ is the Hamming distance between $\pi_i$ and $\pi_j$.*

*Proof.* Let $length(\pi_i) = length(\pi_j) = n$. A mixed-control $Tof_n$ gate will be active to an element if all positive and negative control lines have input bits of 1 and 0 correspondingly. It will invert the target bit . Hence a $Tof_n$ gate can perform a transposition on two elements with Hamming distance 1.

Table 1: Swaps performed by bubble sort for permutation $\pi = (0, 1, 2, 3, 7, 4, 6, 5)$

| (i) | (ii) | (iii) | (iv) | (v) |
|---|---|---|---|---|
| cba | cba | cba | cba | cba |
| 000 | 000 | 000 | 000 | 000 |
| 001 | 001 | 001 | 001 | 001 |
| 010 | 010 | 010 | 010 | 010 |
| 011 | 011 | 011 | 011 | 011 |
| 111 | 100 | 100 | 100 | 100 |
| 100 | 111 | 110 | 110 | 101 |
| 110 | 110 | 111 | 101 | 110 |
| 101 | 101 | 101 | 111 | 111 |

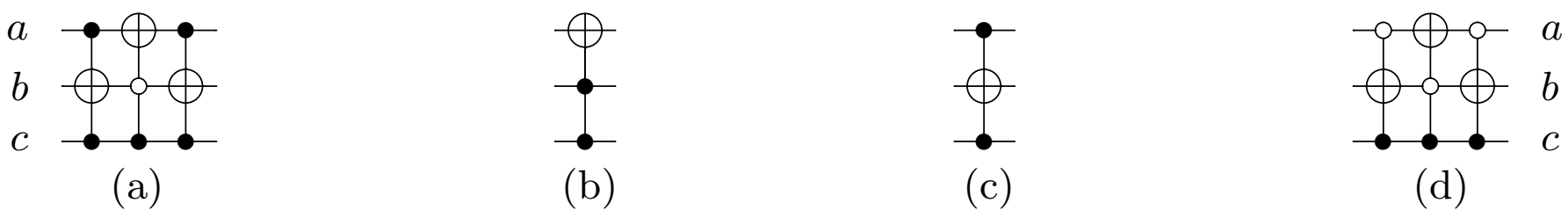

Fig. 1: Reversible Logic Synthesis via Bubble sort

let $\pi_i$ and $\pi_j$ be two elements we want to swap and $h = \delta(\pi_i, \pi_j)$. First we transform $\pi_i$ to $\pi_j$ by applying a series of $Tof_n$ gates. We identify the bits of $\pi_i$ that differ from corresponding bits of $\pi_j$. For each bit, we apply a multi-polarity $Tof_n$ gate whose target line maps to that bit and the control lines corresponds to other bits of $\pi_i$. After each step, Hamming distance will decrease by 1. After $h$-th step, the combined circuit will transform $\pi_i$ to $\pi_j$. The $h$-th $Tof_n$ gate will also invert a bit of $\pi_j$ as Hamming distance between $\pi_j$ and intermediate output of $\pi_i$ is 1. If we apply these series of $Tof_n$ gates in reverse order it will transform $\pi_j$ to $\pi_i$. As there is one common gate, total number of gates needed are $2h - 1$.

As we are applying each $Tof_n$ gate twice, this will cancel out the introduced changes in the permutation elements other than $\pi_i$ and $\pi_j$. □

With this background, one may apply any kind of sorting algorithm to achieve reversible logic synthesis, as the sorting algorithm essentially performs a series of transpositions. In contrast to the previous works, we do not limit the sorting to a series of 3-cycles or transposition of two bitstrings with Hamming distance of 1. The exact steps of the sorting dictate the reversible logic gates. We illustrate this with the help of bubble sort.

### 3.1 Illustrative Example

Let us consider the following permutation, $\pi = (0, 1, 2, 3, 7, 4, 6, 5)$. Bubble sort works by iterating down a list to be sorted from the first element to the last, comparing each pair of elements and switching their positions if necessary. This process is repeated until the list is sorted. Column $i$ and $v$ of table 1 shows the binary representation of $\pi$ and the identity permutation. Column $ii$, $iii$ and $iv$ shows the intermediate output of the bubble sort. The reversible circuits for all steps are shown in Fig. 1 .

Bubble sort belongs to the general class of sorting algorithms referred to as *comparison sort.* Under restrictive conditions sorting can be done without comparisons by integer sort, e.g. radix sort, counting sort. For our problem radix sorting algorithms provide a clear match with improved performance guarantee.

### 3.2 Radix Sort

Among different variants of radix sort, our synthesis method is inspired by the principle of radix exchange sort. Radix exchange sort considers the structure of the keys. Keys are represented in a base M number system ($M = radix$). If M = 2, sorting is done by comparing bits of the binary keys in the same position.An example of a single step of this sort is shown in Fig. 2.

For a $2^n$-element permutation, the efficiency of radix sort is $O(n \cdot 2^n)$, assuming each element is $d = n$ digit. In that sense, it does not offer any improvement over comparison-based sorting. However, the digit count remains relatively small for even a larger permutation set. Further, this allows an alternative treatment of the reversible logic synthesis, as we will see in the section 4. It is interesting to note that the worst-case circuit complexity for an $n$-variable Boolean function, when MMD is applied, is derived to be $(n - 1)2^n + 1$ [15].

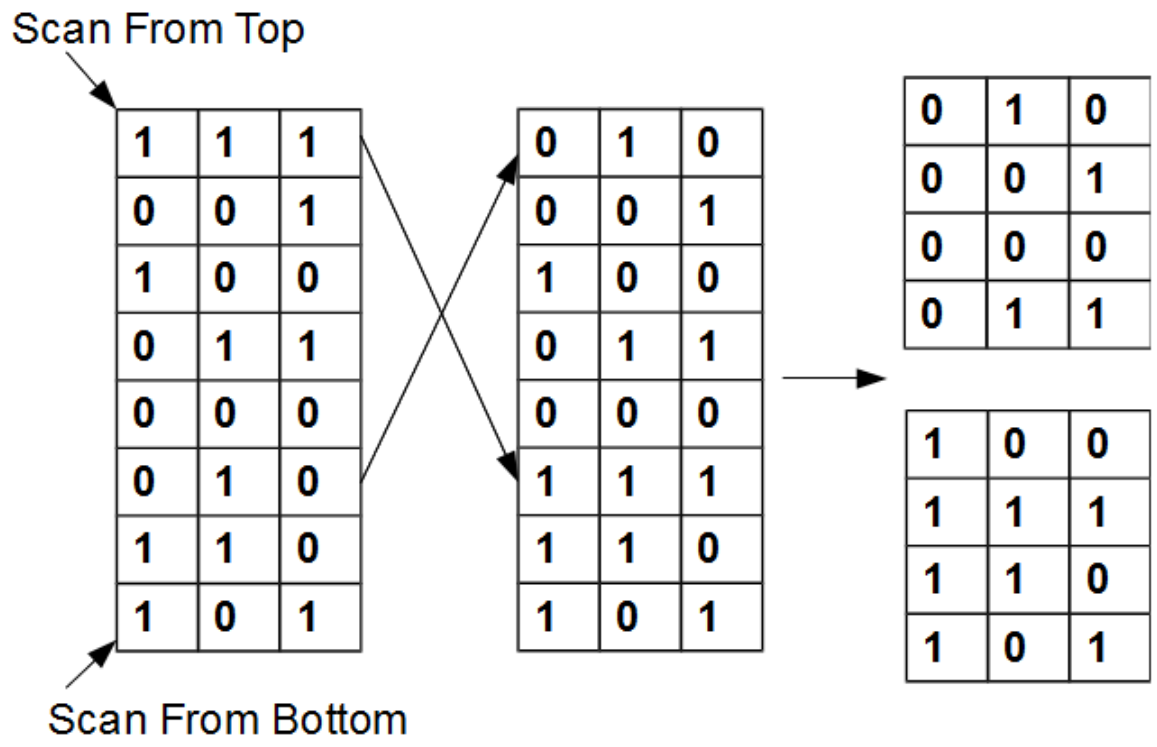

Fig. 2: Radix Exchange Sort

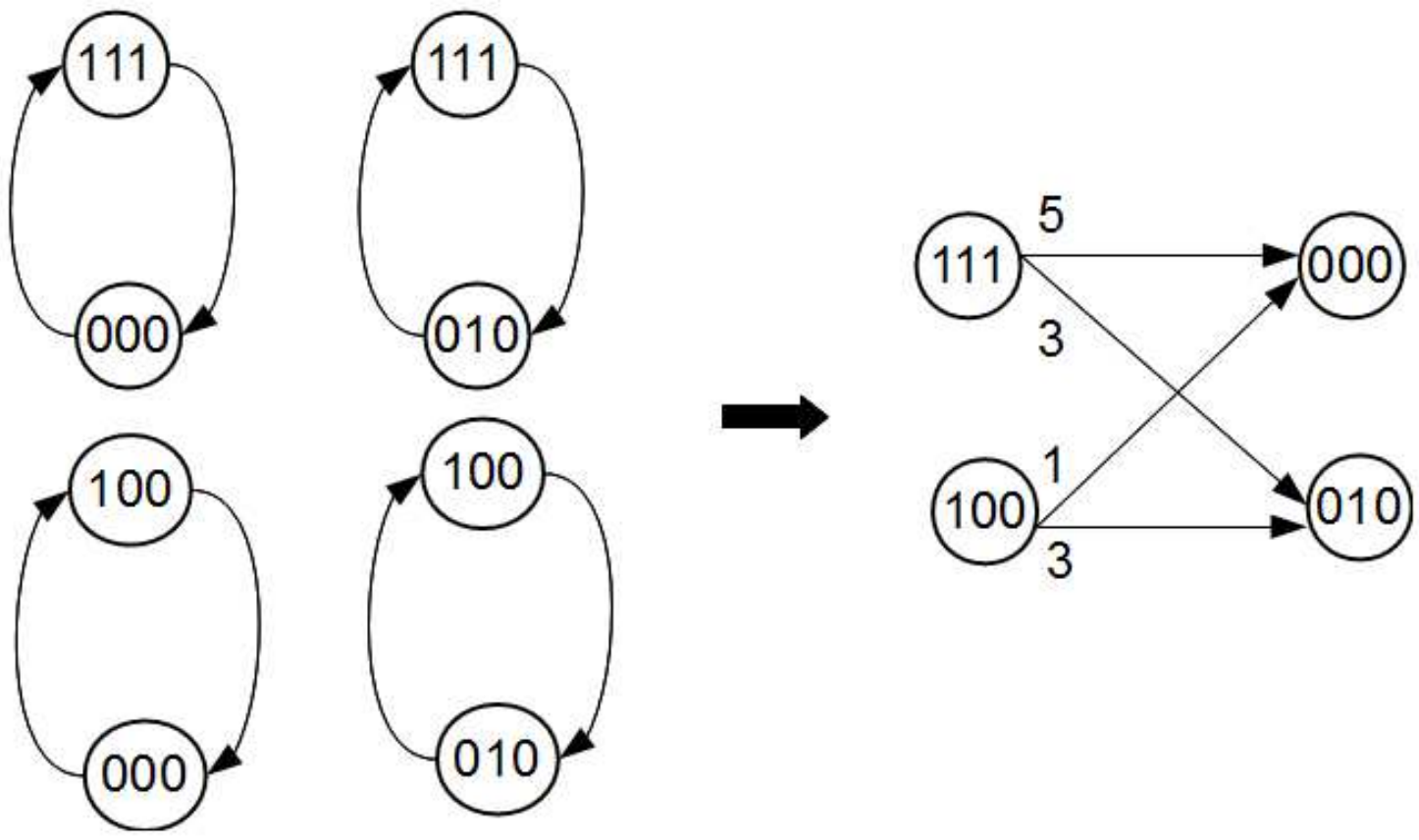

Fig. 3: Functional digraphs and bipartite graph

### 3.3 Functional Digraph to Minimum Weight Perfect Bipartite matching

In radix exchange sort, while swapping several elements, different combinations of transpositions can be used. In reversible synthesis, each transposition incurs a cost. One can optimize the circuit cost by selecting pairs for transposition in the following way.

One can merge the corresponding functional digraphs of all the possible transpositions. In the combined graph, if one replaces each 2-cycle with an undirected edge and assign the cost of transposition as edge weight, the graph becomes a weighted complete bipartite graph. Finding a minimum weight perfect matching in this graph yields desired decomposition with optimized circuit cost.

As an example, for exchanges in Fig. 2 all possible functional digraphs and the complete weighted bipartite graph are shown in Fig. 3. Minimum weight complete matching is $\{(111, 010), (100, 000)\}$. These transpositions will result in a circuit with minimum gate count.

## 4 RevCol: Column-wise Reversible Logic Synthesis

With the background described, a novel reversible logic synthesis algorithm is proposed. Simple application of radix sort with $radix = 2$ shows improvement over the transformation-based algorithm presented in [15]. We first explain with an exemplary permutation network, the working principle of the algorithm. This also serves as a motivational example, where improvement in gate count compared to MMD can be observed. Since the reversible logic synthesis corresponding to radix sort proceeds in a column-wise manner, the algorithm is termed as **RevCol**.

### 4.1 Motivational Example

**RevCol** progresses by synthesizing Quantum circuit that transforms columns to the output specifications one by one at every step. As different orders of input column matching leads to different circuits, **RevCol** constructs circuits for all permutations of column order and selects the optimal circuit. For 3-variable reversible Boolean function 3_17 we shall describe only the executions that leads to the optimal circuit by **RevCol**. The specification is given in table 2. The algorithm matches column a with the corresponding output in the first step. In column $a$, input 111 and 010 does not match with corresponding values in column $a$ of the output. It constructs a complete bipartite graph from these numbers $G(V_1, V_2, E, W)$ where $V_1$ and $V_2$ contains $\{010\}$(0 in column a) and $\{111\}$(1 in column a) respectively and W is set of edge weights where an edge weight is $(2 \cdot (Hammingdistance) - 1)$. On graph $G$, it computes the minimum weight complete matching $(010, 111)$. For swapping 010 and 111, it constructs necessary circuit. After applying this circuit to the input specification, column $a$ of the intermediate output matches with final output. **RevCol** creates two reversible function specification with column $b$ and $c$ such that in each group value of $a$ is same. It recursively constructs circuits for these two functions. When the algorithm returns from recursive calls, it adds positive or negative control lines to the returned circuits based on the value of $a$. The complete circuit is shown in Fig. 4.

| cba | c'b'a' |
|---|---|
| 111 | 000 |
| 110 | 001 |
| 101 | 010 |
| 100 | 011 |
| 011 | 100 |
| 010 | 101 |
| 001 | 110 |
| 000 | 111 |

Table 2: 3_17 Specification

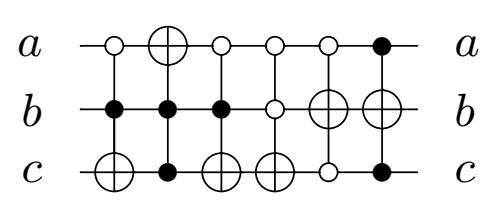

Fig. 4: Circuit for 3_17

### 4.2 Algorithmic Flow

The Algorithm **RevCol** is described by a flowchart in Fig. 5. Minimum weight perfect matching in bipartite graphs is an well-studied problem in theoretical computer science. In our implementation, we used the Hungarian algorithm [20]. Further, several optimizations to improve the efficiency of the synthesized circuit are described. Finally, the runtime complexity of RevCol is analyzed.

### 4.3 Partial Match Optimization

After matching a column, the basic algorithm naively calls the two subgroups recursively to synthesize circuits for them. Sometimes the function specification may be same for both groups. Instead of using two different circuits for two groups, it will be sufficient to use one circuit for both, as shown in Fig. 6 and Fig. 7 for permutation $(7, 5, 3, 1, 6, 4, 2, 0)$.

### 4.4 Optimization via Swap Gates

For 2-bit reversible function specification, sometimes the output is equivalent to swapped columns of input i.e. if input is $(a, b)$, output is $(b, a)$. Instead of using several Toffoli gates, we can use a single swap gate to synthesize the circuit. Fig. 8 show the circuits without and with this optimization for a permutation $(00, 10, 01, 11)$.

### 4.5 Inverted Column Optimization

The size of the circuit is dependent on the number of swaps. For $n$ bits input, if the number of pairs of elements for swapping in greater than $2^{n-2}$, applying a $Tof_1$ gate will result in decreasing the number less than $2^{n-2}$. Let us consider a permutation $\pi(abc) = (110, 111, 100, 010, 101, 011, 000, 001)$. When the naive algorithm first tries to match column $b$ with the corresponding output, it synthesizes circuits to swap $\{100, 110\}$, $\{000, 111\}$ and $\{001, 011\}$. The complete circuit with $(b, a, c)$ as matching column order is shown in Fig. 9. It's size is 13. With this optimization, the algorithm uses $Tof_1$ invert column $b$ and synthesizes circuits to swap $\{000, 111\}$. The optimized circuit (Fig. 10) is of size 8.

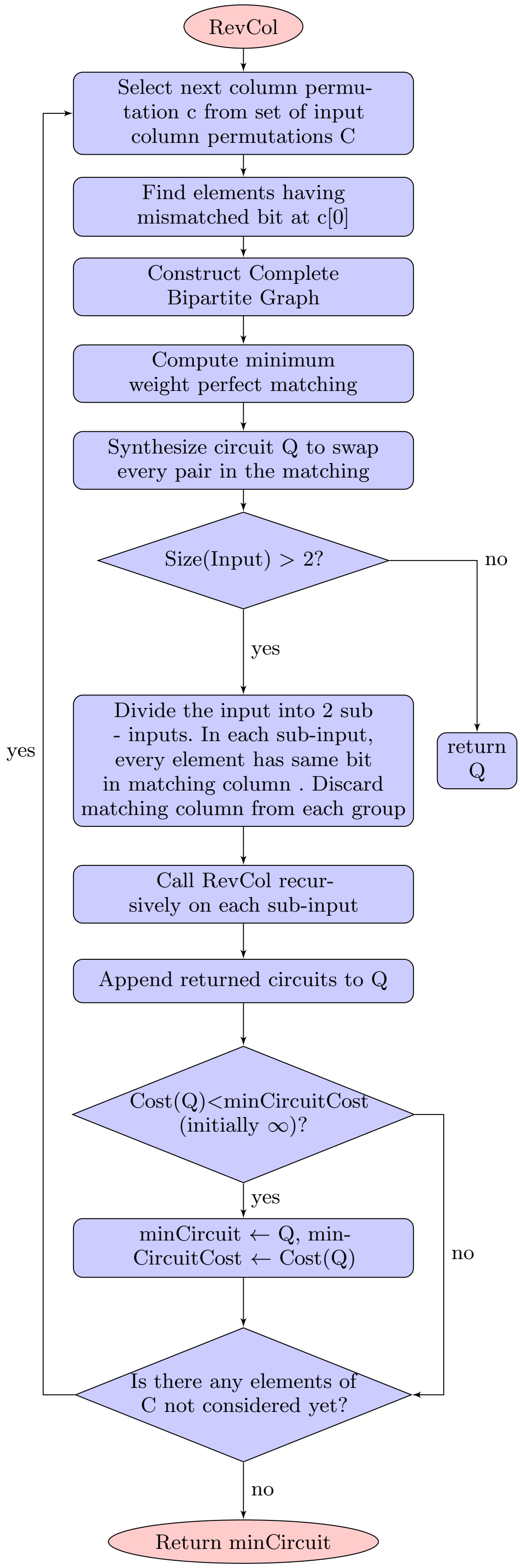

Fig. 5: RevCol flowchart

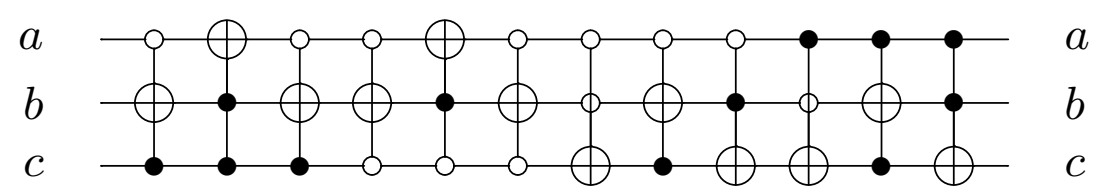

Fig. 6: Circuit without partial matching for permutation $(7, 5, 3, 1, 6, 4, 2, 0)$

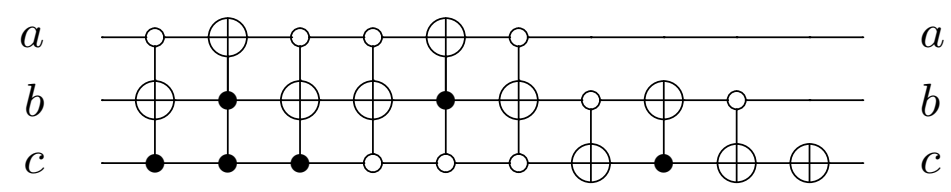

Fig. 7: Circuit with partial matching for permutation $(7, 5, 3, 1, 6, 4, 2, 0)$

### 4.6 Output Permutation

The naive algorithm matches $i$-th column of the input to the $i$-th column of the output. We can also match $i$-th column of the input to the $j$-th column of the output, and also in the other way, and use a swap gate to swap column $i$ and $j$. let $\pi(abc) = (000, 010, 100, 011, 110, 101, 001, 111)$ be a permutation. Using output optimization, the algorithm matches columns $a$ and $b$ to columns $b$ and $a$ of output. Then it uses a SWAP gate to bring the columns to the correct position. The circuits are shown in Fig. 11 .

### 4.7 Transposition Optimization

We discuss an improved way to construct a reversible circuit for a transposition. Initially we used fully controlled Tofolli gates, which results in higher quantum cost. With the following reversible circuit synthesis technique for a transposition, we can construct circuits with less quantum cost. Given a transposition $i_1$, $i_1$, $\ldots i_n \leftrightarrow o_1$, $o_1$, $\ldots o_n$ we can find a reversible circuit that realizes the transposition as follows:

– Fix some $t$ such that $i_t \neq o_t$.
– The circuit consists of three parts $ABA$.
– Part $A$ has CNOT gates for all $j \neq t$ such that $i_j \neq o_j$ with a control of polarity $i_t$ at line $t$ and a target on $j$.
– Part $B$ has one fully controlled Tofolli gate with target at line $t$ and controls according to $o_j$ with $j \neq t$

An example circuit for the transposition $1010 \leftrightarrow 0100$ is shown in Fig. 12. here $t = 3$.

### 4.8 Worst Case Complexity Analysis

For $n$ bits input, there are $2^n$ elements in the permutation. If all elements need to be swapped, we can use a single **NOT** gate to swap them all. In the next case, the number of pairs of elements for swapping is greater

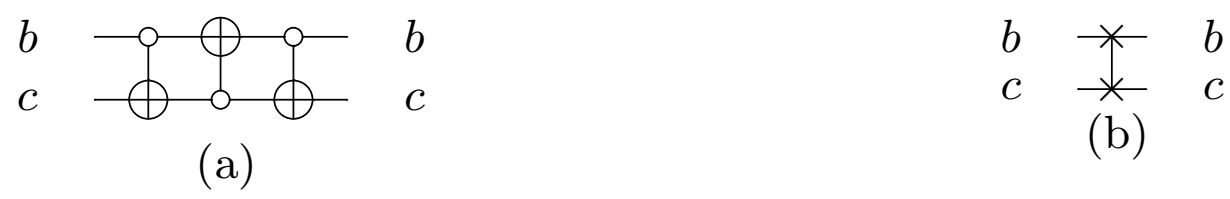

Fig. 8: Circuit without and with swap gates for permutation $(00, 10, 01, 11)$

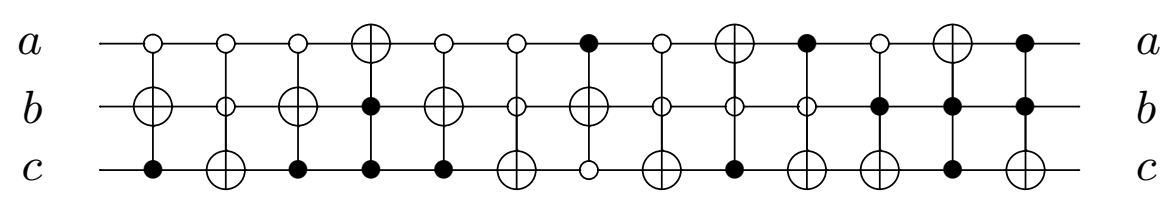

Fig. 9: Circuit without inverted column for permutation $(110, 111, 100, 010, 101, 011, 000, 001)$

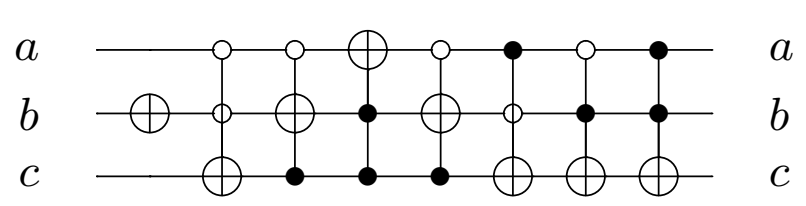

Fig. 10: Circuit with inverted column for permutation $(110, 111, 100, 010, 101, 011, 000, 001)$

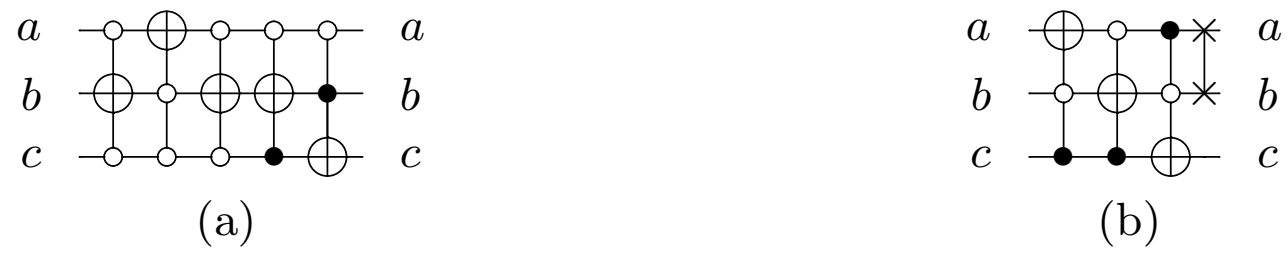

Fig. 11: Circuit without and with output permutation for permutation $(000, 010, 100, 011, 110, 101, 001, 111)$

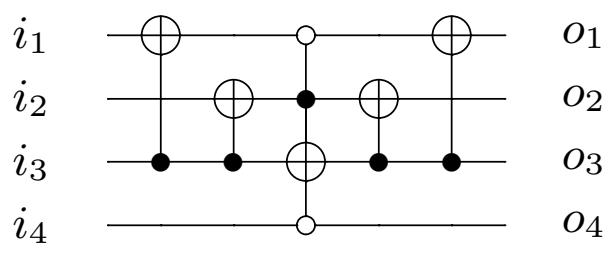

Fig. 12: Circuit for Transposition $1010 \leftrightarrow 0100$

than $2^{n-2}$, but less than $2^{n-1}$. Again we use a **NOT** gate and the number of pairs of elements for swapping becomes at most $2^{n-2}$. Because of inversion of column and matching, the worst case Hamming distance between two elements is $n-1$. Hence, the number of gates in the first step is $1 + 2^{n-2}(2(n-1)-1)$. The total number of gates is:

$1 + 2^{n-2}(2(n-1)-1) + 2/2{+}2 * 2^{n-3}(2(n-2)-1) + 4/2{+}2^2 * 2^{n-4}(2(n-3)-1) + \ldots + n * 1$
$= 1/2 + 1/2(1 + 2 + 2^2 + \ldots + 2^{n-1}){+}2^{n-2}\{2n(n-1)/2 - n\}$
$= 1/2 + 1/2(2^n - 1) + 2^{n-2}(n^2 - 2n)$
$= 2^{n-1} + 2^{n-2}(n^2 - 2n)$
$= 2^{n-2}(n^2 - 2n + 2)$

Hence, the upper bound in terms of gate count for **RevCol** is $2^{n-2}(n^2 - 2n + 2)$. It provides a better result than MMD for $n = 3$ and 4.

### 4.9 Hybrid Algorithm

There are some limitations with RevCol that restricts it's applicability. First, it uses too long control lines that tends to make the Quantum Cost high. Second, it depends on the truth-table manipulations for input and output permutation limiting its scalability. Still, it is competitive compared to MMD in terms of logical depth and hence, might be a good candidate for hybrid synthesis methods. We address both the limitations by exploring a hybrid synthesis method incorporating the synthesis principle of RevCol. In principle, RevCol can be combined with any other scalable synthesis method in the same manner.

**RevCol** is a recursive algorithm. After synthesizing a circuit to convert a particular column of the input function, **RevCol** constructs two reversible functions with one bit less than previous one and call itself recursively. In the hybrid algorithm, for the smaller reversible functions, we synthesize circuits using different reversible methods and select the sub-circuit with optimal cost. Fig. 13 shows a logical flow of the hybrid method. We used RevKit [21] toolkit to implement the hybrid algorithm. From RevKit, we incorporated Reed Muller Synthesis [22], Transformation based Synthesis [15] and Young subgroup based synthesis[23] to the hybrid algorithm.

## 5 Experiments and Benchmarking

We benchmark the efficiency of **RevCol** and the hybrid algorithm by comparing with published results from other synthesis methods [8, 15, 24]. We implemented the synthesis algorithms in C++ using RevKit toolkit. All experiments have been carried out on an Intel(R) Core(TM) i3 CPU with 4 GB of main memory in Linux environment.

Table 3 shows the gate counts for all 3-bit reversible functions(8! = 40320) and compare them to MMD without template matching, and optimal gate count for NCTF multi polarity gate library(N - NOT, C - CNOT, T - Toffoli, F - Fredkin) presented in [25]. **RevCol** with partial match and swap gate optimization provides more optimized circuits than MMD without template matching. Partial match and Output permutation provides much improvement over the naive algorithm. Due to unavailability of a definitive template-set, the template matching is not applied. It is likely that our reported results can improve further with that.

In table 4, we benchmark the results with several random benchmark functions with 4 or 5 variables presented in [26]. The average improvement of Logical Depth is 15% and maximum improvement is 54%.

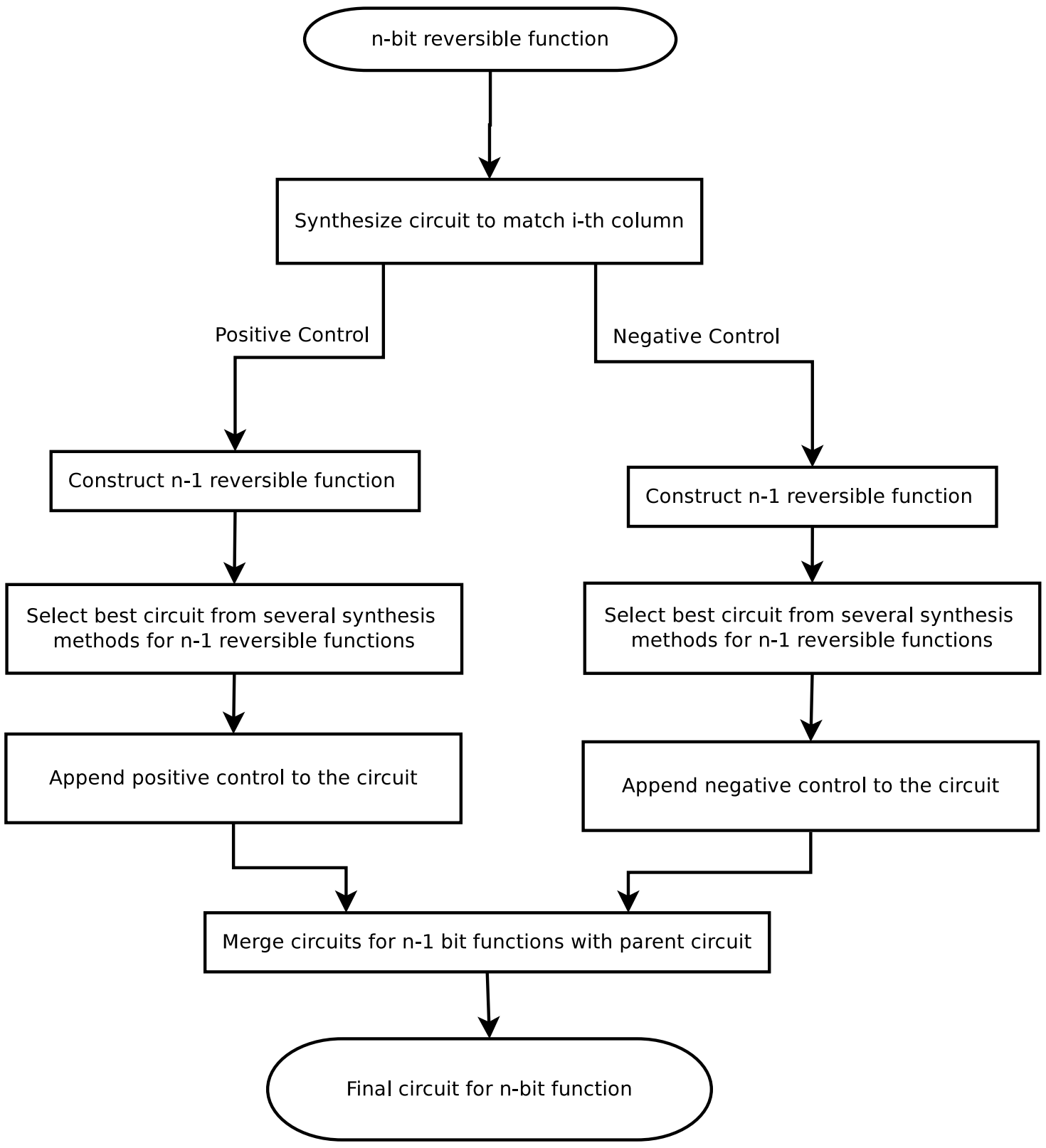

Fig. 13: Hybrid Algorithm

Table 3: No. of Functions with Gate Count for all 3-bit reversible functions

| Gate Count | (a) | (b) | (c) | (d) | (e) | (f) | (g) |
|---|---|---|---|---|---|---|---|
| 16 | 8 | | | | | | |
| 15 | 48 | | | | | | |
| 14 | 72 | | | | | | |
| 13 | 218 | | | | | | |
| 12 | 548 | 14 | 14 | | | 3 | |
| 11 | 1658 | 298 | 266 | | | 86 | |
| 10 | 3528 | 1366 | 1146 | 2 | 6 | 493 | |
| 9 | 6007 | 4108 | 3358 | 414 | 185 | 2312 | |
| 8 | 7964 | 6920 | 6132 | 2648 | 1339 | 6944 | |
| 7 | 7748 | 9680 | 9442 | 7318 | 5982 | 11206 | |
| 6 | 6076 | 7834 | 8262 | 11534 | 12292 | 10169 | 364 |
| 5 | 3895 | 5596 | 6208 | 10282 | 11730 | 5945 | 14175 |
| 4 | 1848 | 3091 | 3692 | 5762 | 6342 | 2375 | 20223 |
| 3 | 572 | 1118 | 1415 | 1929 | 2013 | 650 | 4980 |
| 2 | 114 | 267 | 348 | 394 | 394 | 121 | 544 |
| 1 | 15 | 27 | 36 | 36 | 36 | 15 | 33 |
| 0 | 1 | 1 | 1 | 1 | 1 | 1 | 1 |
| time | 149 | 146 | 125 | 860 | 843 | | |
| Avg | 7.49 | 6.66 | 6.46 | 5.62 | 5.43 | 6.53 | 4.22 |

*(a): naive algorithm*
*(b):(a) plus partial match*
*(c):(b) plus swap gates*
*(d):(c) plus output permutation*
*(e):(d) plus inverted column*
*(f): MMD without template matching*
*(g): optimal (NCTF +/-)*

RevCol synthesizes gates with less logical depth but their QC is higher. [3] The reason is that in RevCol control lines for already sorted columns are often necessary. RevCol employs column by column matching. Once a column is matched, it synthesizes two sub circuits. For a given input, the control lines for already matched columns determine which circuit will be activated for it. It leads to higher number of control lines than MMD and higher QC.

Table 4: Comparison of RevCol and MMD for reversible functions with 4 and 5 variables presented in [26]

| Function | MMD | | RevCol | | % Cost change | |
|---|---|---|---|---|---|---|
| | Gate Count | QC | Gate Count | QC | LD | QC |
| rand4_1 | 19 | 91 | 17 | 194 | -10.5 | 113.2 |
| rand4_2 | 19 | 75 | 14 | 149 | -26.3 | 98.7 |
| rand4_3 | 14 | 62 | 14 | 182 | 0 | 193.5 |
| rand4_4 | 24 | 100 | 11 | 113 | -54.2 | 13 |
| rand4_5 | 19 | 91 | 17 | 172 | -10.5 | 89 |
| rand4_6 | 21 | 93 | 15 | 142 | -28.6 | 52.7 |
| rand4_7 | 15 | 79 | 17 | 209 | 13.3 | 164.6 |
| rand4_8 | 15 | 55 | 13 | 105 | -13.3 | 90.0 |
| rand5_1 | 53 | 437 | 41 | 1087 | -22.6 | 148.7 |
| rand5_2 | 50 | 390 | 40 | 903 | -20 | 131.5 |
| rand5_3 | 51 | 480 | 45 | 1053 | -11.8 | 119.4 |
| rand5_4 | 48 | 384 | 46 | 1166 | -4.2 | 203.6 |
| rand5_5 | 50 | 458 | 43 | 1155 | -14 | 152.2 |
| rand5_6 | 48 | 440 | 42 | 1038 | -12.5 | 135.9 |
| rand5_7 | 51 | 384 | 45 | 1093 | -11.8 | 184.6 |
| rand5_8 | 51 | 396 | 44 | 1105 | -13.7 | 179 |
| Average | 34.2 | 250.9 | 29 | 616.6 | -15 | 129.4 |

In table 5, we show the results of the benchmark for some 4-bit reversible functions previously appeared in the literature. We compared the gate count against the optimal gate count proposed in [24]. We did not compare the QC as it was not present in [24]. As we can see, RevCol normally synthesizes circuit with higher gates than optimal circuits. Hybrid algorithm constructs circuits that have same or one more gates as the optimal circuits .

In table 6, we show the results of the benchmark for some reversible functions comparing against [11]. Here we compare the gate costs and quantum costs. BDD generally appends additional ancilla to the circuit for synthesis. In all cases, Hybrid algorithm has less number of lines. The gate cost and the quantum cost both are less for the Hybrid algorithm compared to the BDD synthesis.

In table 7, we show the results for the Hybrid algorithm without input and output permutation for some larger reversible functions comparing against [11]. Here we can see, without the input and output permutation, the hybrid algorithm performs worse than BDD based synthesis. The table also shows the synthesis method used by the hybrid algorithm to achieve the best result. The low cost in some cases can be attributed to the pattern of the reversible functions and selection of synthesis in Hybrid method.

In table 8, we compared the results for the Hybrid algorithm without input and output permutation with that from Ancilla free BDD synthesis algorithm [28]. The table shows that RevCol Hybrid achieves better quantum cost but the gate cost is higher. The lower cost for *cycle*10_2 can be attributed to its repetitive pattern in the function.

## 6 Conclusion and Future Work

In this paper, we introduced a novel synthesis approach by realizing the principles of sorting. Any reversible function inherently performs a sorting. Hence by using the principles of radix sort, we proposed a new algorithm. The experimental results show that 54% improvement in logical depth over MMD can be achieved with our algorithm. Based on the principles of RevCol, we described a scalable hybrid synthesis method.

In future, we will explore the theoretical connection and practical derivation of other sorting algorithms with reversible logic synthesis

[3] For calculating QC, we used the metrics presented in [27] based on the work of [5]

Table 5: Benchmark for 4-bit reversible functions against optimal gate count[24]

| Name | Specification | Optimal Circuit GC[24] | RevCol GC | Hybrid GC |
|---|---|---|---|---|
| 4bit-7-8 | 0,1,2,3,4,5,6,8,7, 9,10,11,12,13,14,15 | 7 | 7 | 7 |
| decode42 | 1,2,4,8,0,3,5,6,7, 9,10,11,12,13,14,15 | 10 | 11 | 10 |
| imark | 4,5,2,14,0,3,6,10, 11,8,15,1,12,13,7,9 | 7 | 9 | 9 |
| mperk | 3,11,2,10,0,7,1,6, 15,8,14,9,13,5,12,4 | 9 | 14 | 10 |
| oc5 | 6,0,12,15,7,1,5,2, 4,10,13,3,11,8,14,9 | 11 | 15 | 12 |
| oc6 | 9,0,2,15,11,6,7,8, 14,3,4,13,5,1,12,10 | 12 | 17 | 14 |
| oc7 | 6,15,9,5,13,12,3, 7,2,10,1,11,0,14,4,8 | 13 | 15 | 14 |
| oc8 | 11,3,9,2,7,13,15, 14,8,1,4,10,0,12,6,5 | 12 | 16 | 13 |
| rd32 | 0,7,6,9,4,11,10, 13,8,15,14,1,12,3,2,5 | 4 | 6 | 4 |
| shift4 | 1,2,3,4,5,6,7,8,9, 10,11,12,13,14,15,0 | 4 | 4 | 4 |
| 4_49 | 15,1,12,3,5,6,8,7, 0,10,13,9,2,4,14,11 | 12 | 17 | 16 |

Table 6: Comparison of RevCol Hybrid and BDD [11] for small functions

| Function | BDD | | | Hybrid | | |
|---|---|---|---|---|---|---|
| | Lines | Gate Count | QC | Lines | Gate Count | QC |
| 3_17_6 | 10 | 20 | 50 | 3 | 5 | 27 |
| miller_5 | 8 | 15 | 38 | 3 | 5 | 9 |
| ham3_28 | 10 | 18 | 46 | 3 | 7 | 25 |
| peres_4 | 24 | 100 | 11 | 3 | 2 | 8 |
| 4_49_7 | 18 | 25 | 114 | 4 | 14 | 94 |
| aj-e11_81 | 19 | 45 | 113 | 4 | 18 | 42 |

Table 7: Comparison of RevCol Hybrid (without input and output permutation) and BDD [11]

| Function | BDD | | | Hybrid | | | |
|---|---|---|---|---|---|---|---|
| | Lines | Gate Count | QC | Lines | Gate Count | QC | Synthesis |
| mod5d2_17 | 19 | 42 | 102 | 5 | 9 | 25 | tbs |
| hwb6_14 | 53 | 167 | 437 | 6 | 251 | 927 | rms |
| graycode6_11 | 16 | 20 | 45 | 6 | 5 | 5 | tbs |
| ham7_29 | 36 | 88 | 224 | 7 | 64 | 1118 | ysg |
| hwb7_15 | 84 | 284 | 744 | 7 | 600 | 3680 | rms |
| hwb8_64 | 129 | 456 | 1195 | 8 | 1538 | 11294 | rms |

Table 8: Comparison of Ancilla free BDD [28] and RevCol Hybrid without input and output permutation

| Function | BDD | | | Hybrid | | | |
|---|---|---|---|---|---|---|---|
| | Lines | Gate Count | QC | Lines | Gate Count | QC | Synthesis |
| urf2_73 | 8 | 268 | 24066 | 8 | 1463 | 11507 | rms |
| urf1_72 | 9 | 563 | 74858 | 9 | 3569 | 38613 | rms |
| hwb9_65 | 9 | 584 | 73465 | 9 | 3558 | 37250 | rms |
| urf3_75 | 10 | 1081 | 162225 | 10 | 6859 | 114643 | rms |
| urf4_89 | 11 | 2641 | 491645 | 11 | 19092 | 355632 | rms |
| cycle10_2 | 12 | 27 | 4200 | 12 | 19 | 6079 | tbs |